\documentclass[aps, twocolumn,floats, showpacs]{revtex4}
\usepackage{amsmath,amssymb,graphicx}
\usepackage{epsfig}
\usepackage{graphicx}
\usepackage{enumerate}

\newcommand{\beq}{\begin{equation}}
\newcommand{\eeq}{\end{equation}}

\newcommand{\bea}{\begin{eqnarray}}
\newcommand{\eea}{\end{eqnarray}}

\begin{document}

\title{
Thermal conduction in molecular chains: Non-Markovian effects}
\author{Dvira Segal}
\affiliation{
Chemical Physics Theory Group, Department of Chemistry,
University of Toronto,
80 St. George Street, Toronto, Ontario M5S 3H6, Canada}

\date{\today}
\begin{abstract}
We study the effect of non-Markovian reservoirs on the heat conduction properties
of short to intermediate size molecular chains.
Using classical molecular dynamics simulations, we show that the distance
dependence of the heat current is determined not only by the molecular properties,
rather it is also critically influenced by the spectral properties of the heat baths,
for both harmonic and anharmonic molecular chains.
For highly correlated reservoirs the current of an anharmonic chain
may exceed the flux of the corresponding harmonic system.
Our numerical results are accompanied by a simple single-mode heat conduction model
that can capture the intricate distance dependence obtained numerically.
\end{abstract}

\pacs{
}

\maketitle

\section{Introduction}

The problem of heat conduction through molecular structures has recently
attracted lot of attention \cite{Kim, Braun, Segalman, RectifE, Dlott, films}
with potential applications in thermal machinery \cite{Tunable,pump},
information processing and computation \cite{Gate,Circuit},
and thermoelectricity \cite{Reddy,Hochbaum}.

One of the major open questions here is what are the factors that dominate thermal
transport, the molecular structure, the contacts, or both \cite{Segalman}?
Another central issue is the determination of the system size ($N$)
dependence of the heat current $J$.
While Fourier's law suggests the relation $J \propto N^{-1}$, extensive studies of heat flow
in low dimensional systems have resolved a $J \propto N^{-\alpha}$ behavior,
where $\alpha$ usually deviates from 1 \cite{Lepri-rev}.
Specifically, for harmonic chains one gets
$\alpha=0$ in the Markovian limit \cite{Lebowitz},
i.e. the heat current does not depend on system size.
This ballistic behavior results from the lack of scattering mechanisms between normal modes.
In the harmonic limit the heat current thus reflects the spectral properties of
the thermal reservoirs. In other words, it
crucially depends on the details of the boundary conditions \cite{Dhar,Saito}.
In contrast, in strongly anharmonic systems where local thermal equilibrium exists,
one expects that the steady state energy current will not depend on
the properties of the contacts.

Recent studies of heat transport in disordered low-dimensional
harmonic chains have manifested the influential
role of the contacts' spectral properties on the asymptotic $\alpha$ value \cite{Dhar,Dhar2D}.
Subsequent works exemplified this effect within anharmonic lattices \cite{Zhao,Barik}.
While these works have typically focused on the  asymptotic length behavior,
our objective here is to systematically study the effect of the reservoirs' spectral properties
on the thermal transport in 
{\it short to intermediate} size molecular junctions that are of experimental relevance.

Using classical molecular dynamics simulations,
we analyze the distance dependence of the current and the chain's temperature profile
for Markovian and non-Markovian thermal baths considering either harmonic or anharmonic
internal molecular interactions.
We find that the spectral properties of the reservoirs
play a crucial role in determining the size dependence of the thermal current.
Thus, one should carefully interpret experimental results \cite{Segalman},
as both molecular structure and the properties of the boundaries
critically determine the junction conductivity.
Another interesting finding is that for highly correlated noise, anharmonic chains
conduct more effectively than the corresponding harmonic systems.
We qualitatively explain our numerical results using 
a single-mode heat conduction model that can be solved analytically \cite{Rectif, NDR}.




%



\section{Molecular dynamics simulations}

We present here detailed classical molecular dynamics simulations of
steady state heat transfer through one-dimensional (1D) molecular chains
coupled to non-Markovian reservoirs.
We model the molecule as a chain of $N$ identical atoms. The end particles 1 and $N$
are connected to heat baths of temperatures $T_L$ and $T_R$ respectively.
The dynamics is governed by the generalized Langevin equation
\bea
\ddot x_k(t)&=&-\frac{1}{m}\frac{\partial H_0}{\partial x_k},  \,\,\, k=2,3.. N-1
\nonumber\\
\ddot x_1(t)&=&-\frac{1}{m}\frac{\partial H_0}{\partial x_1}  -\int_0^t dt'
\gamma_L(t-t')\dot x_1(t') + \eta_L(t),
\nonumber\\
\ddot x_N(t)&=&-\frac{1}{m}\frac{\partial H_0}{\partial x_N}  -\int_0^t dt'
\gamma_R(t-t')\dot x_N(t') + \eta_R(t).
\nonumber\\
\label{eq:EOM}
\eea
$x_k$ is the 
position of the $k$ particle of mass $m$, and $p_k$ [see Eq. (\ref{eq:EOM2})] is the particle
momentum.
$H_0$ is the internal molecular Hamiltonian.
$\gamma_L$ and $\gamma_R$ are friction constants and $\eta_L$
and $\eta_R$ are fluctuating forces that represent the effect of the
thermal reservoirs. These terms are related through the
fluctuation-dissipation relation ($n=L,R$)
\bea
\langle \eta_n \rangle=0 ; \,\,\,\,
\langle \eta_n(t) \eta_n(t')\rangle=\frac{k_B T_n}{m}\gamma_n(t-t'),
\eea
where $k_B$ is the Boltzmann constant.
We consider here an exponentially correlated Ornstein-Uhlenbeck (O-U) noise \cite{OU}
\bea
\gamma_n(t-t')=\frac{\epsilon_n}{\tau_c^n}e^{-|t-t'|/\tau_c^n},
\eea
with the intensity $\epsilon$  and a  correlation time $\tau_c$.
For short correlation times the heat baths
generate an uncorrelated (white) noise,
$\gamma_n(t-t')\xrightarrow{\tau_c^n \rightarrow 0} 2\epsilon_n \delta(t-t')$.
The Fourier transform of the O-U correlation function, to be used below, is
\bea
\gamma_n(\omega)\equiv \int e^{-i\omega t} \gamma_n(t) dt = \frac{2\epsilon_n}{1+(\omega\tau_c^n)^2}.
\label{eq:gafo}
\eea
A simple  approach for implementing the O-U noise in numerical simulations
is to introduce auxiliary dynamical variables $y_1(t)$  and $y_{N}(t)$ for the
$L$ and $R$ baths respectively \cite{Luczka}.
The new equations of motion for the first particle  are
\bea
\dot x_1(t) &=&\frac{p_1(t)}{m}
\nonumber\\
\ddot x_1(t)&=&-\frac{1}{m}\frac{\partial H_0}{\partial x_1} -y_1(t) +\eta_L(t)
\nonumber\\
\dot y_1(t)&=&-\frac{y_1(t)}{\tau_c^L} +\frac{\epsilon_L}{m\tau_c^L}p_1(t)
\nonumber\\
\dot \eta_L(t)&=&-\frac{\eta_L(t)}{\tau_c^L} + \frac{1}{\tau_c^L}\sqrt{\frac{2\epsilon_Lk_B T_L}{m}}\mu_L(t),
\label{eq:EOM2}
\eea
where $\mu_L(t)$ is a Gaussian white noise,
$\langle \mu_L(t)\rangle=0$ and $\langle \mu_L(t) \mu_L(t')\rangle= \delta(t-t')$.
An equivalent set of equations exists for the $N$ particle, interacting with the $R$ thermal bath.
The coupled equations, Eq. (\ref{eq:EOM}) for particles 2..$N-1$ and 
(\ref{eq:EOM2}) with its $N$ equivalent,
are integrated using the fourth order Runge-Kutta method
to yield the positions and velocities of all particles.
The heat flux can be calculated from the trajectory using \cite{Lepri-rev}
\bea
J=\frac{1}{2(N-1)} \sum_{k=1}^{N-1} \langle (v_k+ v_{k+1})  F(x_{k+1}-x_k)  \rangle,
\eea
where $F(r)= -dH_0(r)/dr$, $v_k=p_k/m$,
and we average over time after steady state is achieved.

We describe next the molecular structure of the chain.
We model the interactions between the atoms
using a Morse potential of dissociation energy $D$,  width $\alpha$,
and an interatomic equilibrium separation $x_{eq}$,
\bea
H_0=\sum_{k=1}^{N}\frac{p_k^2}{2m}+
D\sum_{k=1}^{N} \left[e^{-\alpha(x_{k+1}-x_k-x_{eq})}-1\right]^2.
\eea
We consider two sets of parameters:
In the first case the potential width is taken
to be very small $\alpha \ll 1$, so as the potential energy is practically harmonic
with a force constant $2D \alpha^2$. We refer to this model as "harmonic".
We also use parameters where the anharmonic coefficient is large,
$\alpha/\sqrt{m D}$ of order 1. We refer to the later case as "anharmonic".

Unless otherwise stated, in the numerical simulation presented below we
have typically used the following parameters: $D= 367.8/\nu^2$ kJ/mol,
$\alpha=1.875 \nu$ \AA$^{-1}$, $x_{eq}=1.54$ \AA, and $m$=12 gr/mole. 
These numbers describe a c-c stretching mode for $\nu=1$ \cite{Lifson}.
We take $\nu=0.01$ for the harmonic model,
while in the anharmonic case we use $\nu=6$.
We also assume that the two reservoirs have the same type of spectral function (O-U)
with equal strength $\epsilon=\epsilon_n$ and noise correlation time $\tau_c=\tau_c^n$.
Depending on the situation, we have used integration time step
$\Delta t = 10^{-3}-10^{-4}$ $1/\omega$,
where $\omega$ is the molecular frequency in the harmonic limit.
We also take care of the required inequality $\Delta t \ll \tau_c$.

\begin{figure}[htbp]
\vspace{0mm} \hspace{0mm}
 {\hbox{\epsfxsize=80mm \epsffile{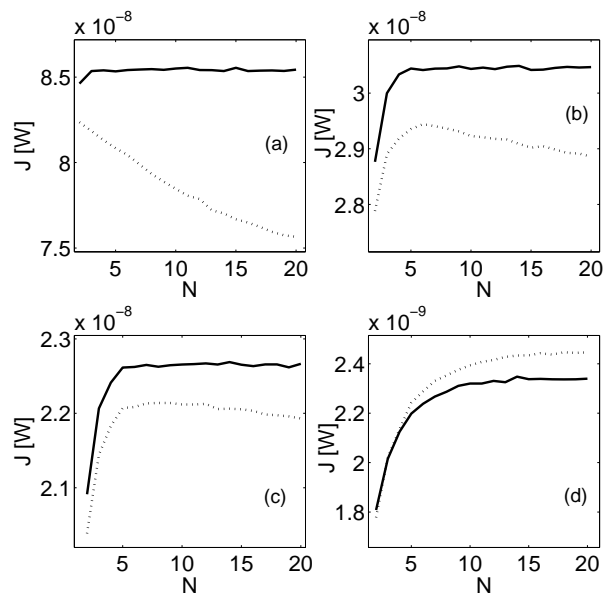}}}
\caption{Distance dependence of the heat current in non-Markovian
systems for harmonic (full), and anharmonic (dotted) models. (a)
Gaussian white noise; (b) O-U noise with $\tau_c$=8 $\times10^{-3}$ ps; (c) O-U
noise with $\tau_c=0.01$ ps; (d) O-U noise with $\tau_c$=0.04 ps.
$T_R=300K$, $T_L=0K$, $\epsilon=50$ ps$^{-1}$
 in all cases. }
\label{Fig4}
\end{figure}

\begin{figure}[htbp]
\vspace{0mm} \hspace{0mm} {\hbox{\epsfxsize=80mm
\epsffile{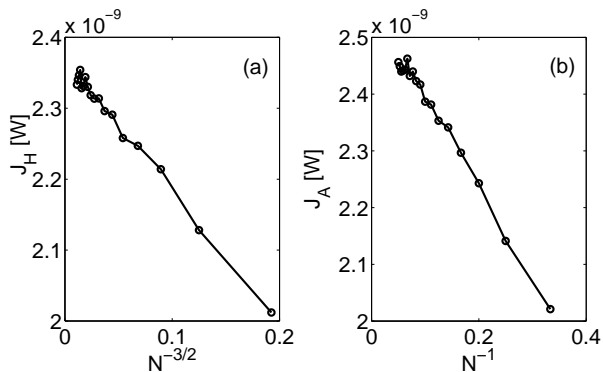}}}
\caption{Resolving the distance
dependence of the thermal current in non-Markovian O-U systems for
harmonic (a) and anharmonic (b) models, $\tau_c$=0.04 ps,
$T_R=300K$, $T_L=0K$, $\epsilon=50$ ps$^{-1}$. }
\label{Fig4a}
\end{figure}

\begin{figure}[ht]
\vspace{0mm} \hspace{0mm} {\hbox{\epsfxsize=75mm
\epsffile{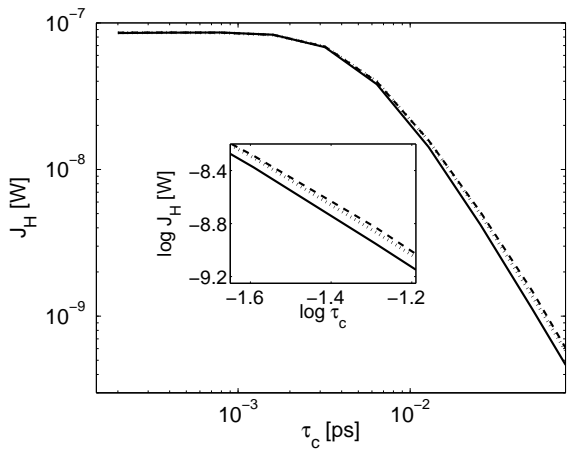}}} \caption{Decrease of heat current with
increasing bath correlation time for harmonic systems. $T_R=300K$,
$T_L=0K$, $\epsilon=50$ ps$^{-1}$. $N$=2 (full), $N=5$ (dotted),
$N=10$ (dashed). The inset zooms on the high $\tau_c$ values. }
\label{Figt1}
\end{figure}

\begin{figure}[ht]
\vspace{0mm} \hspace{0mm} {\hbox{\epsfxsize=75mm \epsffile{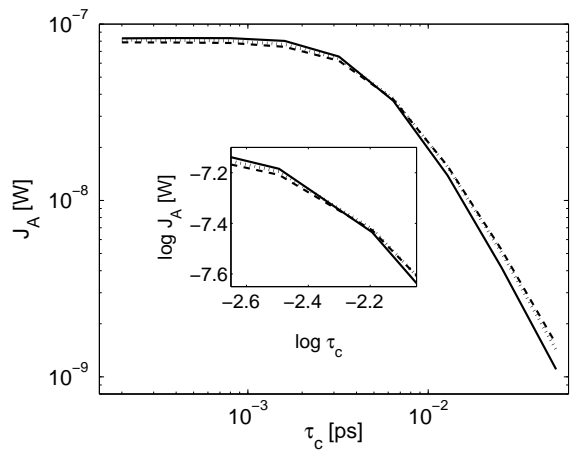}}}
\caption{Decrease of heat current with increasing bath correlation time for an anharmonic molecular model.
$N$=2 (full), $N=5$ (dotted), $N=10$ (dashed). $T_R=300K$, $T_L=0K$,
$\epsilon_n=50$ ps$^{-1}$. The inset zooms on intermediate $\tau_c$
values where $J_A$ is independent of length.} \label{Figt2}
\end{figure}


Figure \ref{Fig4} presents the heat current for harmonic ($J_H$) and anharmonic ($J_A$) chains
calculated with different memory times $\tau_c$.
Panel (a) shows the heat current in the Markovian limit.
We find that the energy flux in harmonic systems does not depend on size,
while it decays with distance for anharmonic chains
in agreement with standard results \cite{Lepri-rev}.
When the noise correlation time is increased, an interesting behavior is observed [panel (b)]:
While $J_H$ remains a constant to a good approximation,
the anharmonic flux manifests an initial rise, followed by a decay for long enough chains.
As the memory time is further increased (c),
the current of an anharmonic system saturates,
and is approximately a constant over the relevant sizes.
This observation interestingly shows a counteracting effect between the molecular contribution
to the heat current and the reservoirs spectral properties.
For highly correlated reservoirs (d) both harmonic and anharmonic currents are slightly enhanced
with distance. Surprisingly, in this case the anharmonic junction conducts better than a fully harmonic system.

We explain next these observations. First we clarify  why  $J_H$- and
$J_A$ for short chains- increase with $N$ for non-Markovian baths.
As was shown in Ref. \cite{heatcond} the dominant heat conducting
vibrational modes of alkane chains are shifted towards lower
frequencies with increasing molecular size. Since within the O-U
model $\gamma(\omega)$ (reflecting the system-bath coupling) 
is larger at lower frequencies, the current gets enhanced with distance.

Next we explain the intricate current-distance behavior of
anharmonic systems. Anharmonic interactions lead to scattering
processes between the molecular modes. These scattering effects are more influential with increasing chain length.
In Markovian systems this
results in the enhancement of the junction resistance, thus it leads to the reduction of 
current with $N$.
However, in non-Markovian systems these
scattering effects are actually beneficial for transferring energy from
molecular modes which are above the reservoirs' cutoff frequencies,
into low energy modes that overlap with the solids vibrations. The
interplay between these two effects leads to a rich behavior: If
$\tau_c^{-1}$ is higher than the molecular frequencies, here of the
order of 150 ps$^{-1}$, anharmonic effects lead to the
reduction of current with size, see panels (a)-(b). In the opposite small cutoff limit
(d), $\tau_c^{-1}= 25$ 1/ps,  harmonic systems can transfer only
those modes that are in the reservoirs energy window, 
while anharmonic junctions better conduct by scattering high energy
modes into low frequencies. For $\tau_c \sim 0.01$ ps the two
effects practically cancel and $J_A$ weakly depends on distance (c).


Fig. \ref{Fig4a} presents the distance dependence of the energy flux 
for both harmonic and anharmonic chains
assuming solids of long memory time, $\tau_c =0.04$ ps.
While it is difficult to make a definite conclusion, we find that $J_H$ and $J_A$ obey
different functional forms.

Next we systematically explore the dependence of the heat flux on the reservoirs' correlation time.
Figs. \ref{Figt1} and \ref{Figt2} manifest that the Markovian behavior
sustains for times up to $\tau_c\sim5 \times 10^{-3}$ ps.
For longer correlation times the heat current significantly decays with $\tau_c$
for both harmonic and anharmonic chains.
The inset of Fig. \ref{Figt2} further shows that for
$\tau_c \sim 6 \times 10^{-3}$ ps the anharmonic heat flux
is practically distance independent, see also  Fig. \ref{Fig4}(c).
As discussed above, this intriguing behavior results from an effective
cancellation between internal molecular interactions, leading to the decay of current with $N$,
and the reservoirs properties, which can lead to an enhancement of current with distance.
In the next section we present a simple analytical model that can capture this intricate behavior.

\begin{figure} [ht]
\vspace{0mm} \hspace{0mm} {\hbox{\epsfxsize=70mm
\epsffile{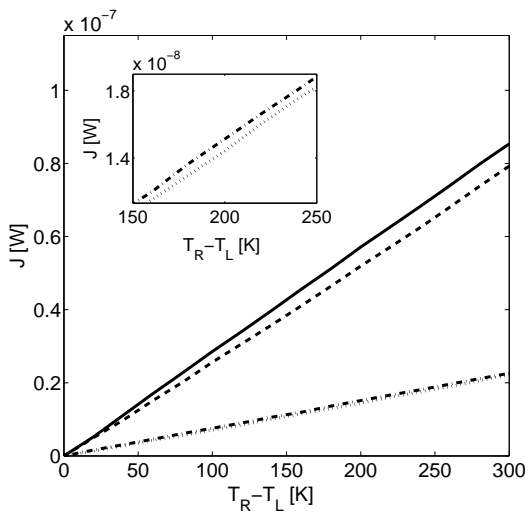}}}
\caption{Temperature dependence of the thermal flux,  $N$=8, $T_R$=300 K.
Harmonic chain with white noise (full);
Anharmonic chain with white noise (dashed);
Harmonic chain with O-U noise, $\tau_c$=0.01 ps, $\epsilon$= 50 ps$^{-1}$
(dashed-dotted);
Anharmonic chain  with O-U noise, $\tau_c$=0.01 ps, $\epsilon$= 50 ps$^{-1}$
(dotted).
The inset zooms on the O-U simulations.
}
\label{FigT}
\end{figure}

In Fig. \ref{FigT} we study the temperature dependence of the heat current 
for both Markovian and non-Markovian
chains for a representative length $N$=8.
We find that both harmonic and anharmonic systems show a linear current-temperature 
relationship in the range $T=0-300$ K.
The thermal conductance ($J/\Delta T$) calculated from Figs. \ref{Fig4}-\ref{Figt2}
is therefore approximately independent of temperature.

\begin{figure} [ht] 
\vspace{0mm} \hspace{0mm} {\hbox{\epsfxsize=70mm
\epsffile{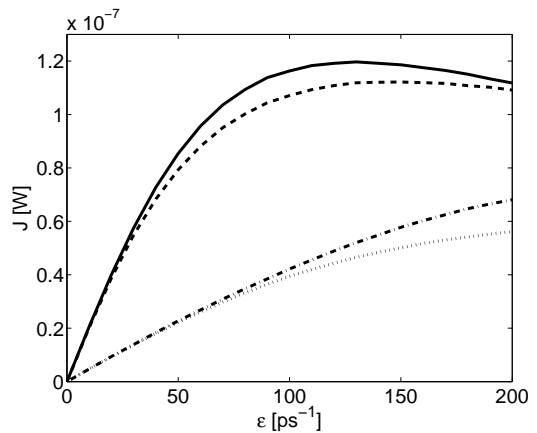}}}
\caption{ 
Thermal flux as a function of system-bath coupling strength $\epsilon$,  
 $T_R$=300 K, $T_L$=0 K. Parameters and lines setting are
the same as in Fig. \ref{FigT}.
}
\label{epsilon}
\end{figure}

We have also analyzed in Fig. \ref{epsilon} 
the $\epsilon$ dependence of the current for Markovian and O-U systems, and found an
approximate linear relation in the low dissipation regime. 
For Markovian baths the current decreases 
as $\epsilon^{-1}$  when the coupling is strong, $\epsilon>100$ ps$^{-1}$. 
We expect that the colored noise model will demonstrate a similar behavior 
for very strong molecule-bath interactions \cite{Barik}.

We conclude this section by noting 
that the qualitative behavior observed above (Figs. \ref{Fig4}-\ref{Figt2}) applies 
for a broad range of temperatures and coupling parameters. We expect that  
similar characteristics will be discovered in molecular systems of various anharmonic internal interactions.


\section{Single-mode heat conduction model}

We present here a simple model that yields a qualitative explanation for the
influential role of non-Markovian reservoirs on the heat transfer
properties of short to intermediate ($N=2-20$) molecular chains.
In our simple picture, heat current in a linear molecular system is dominated by a
specific vibrational mode of frequency $\omega_0$.
The total Hamiltonian includes three terms, $H=H_B+H_0+H_I$, where 
for a harmonic local mode
\bea
H_B&=& \sum_{j,n} \frac{p_{j,n}^2}{2m_{j}} + \frac{1}{2}m_{j} \omega_{j}^2q_{j,n}^2
\nonumber\\
H_0&=&  \frac{p_0^2}{2M} + \frac{M\omega_0^2q_0^2 }{2}
\nonumber\\
H_I&=& \sum_{j,n} \lambda_{j,n} q_{j,n} q_0.
\label{eq:H}
\eea
$H_B$ includes two thermal reservoirs $n=L,R$ of different temperatures, each consisting a set
of independent harmonic oscillators with masses $m_j$, coordinates $q_{j,n}$, and momenta $p_{j,n}$.
$H_0$ represents the (single) relevant molecular mode with
coordinate $q_0$, momentum $p_0$, mass $M$ and frequency $\omega_0$.
It can be written equivalently in the energy representation as
$H_0=\sum_{l=0,1,2..} l \omega_0 |l \rangle \langle l|$;  $\hbar\equiv 1$.
If we sum over states up to infinity ($l\omega_0\gg T_n$), this term describes a harmonic mode
as in Eq. (\ref{eq:H}). We can also model an anharmonic molecule by truncating the 
single mode spectrum  to include only few vibrational states \cite{Rectif,NDR}.
The system-bath interaction is taken to be bilinear, with $\lambda_{j,n}$ as the coupling constant.

Assuming weak molecule-bath coupling at both ends,
an analytical expression for the heat current in steady state can
be derived using the master equation formalism \cite{Rectif,NDR}.
In the harmonic limit and at high temperatures ($T>\omega_0$)
the thermal current is given by a Landauer type expression \cite{Rectif} 
\bea
J_H=\frac{\gamma_L(\omega_0) \gamma_R(\omega_0)}{\gamma_L(\omega_0)+\gamma_R(\omega_0)} k_B \Delta T,
\label{eq:JH}
\eea
where $\Delta T=T_R-T_L$ and
$\gamma_n(\omega)= \frac{\pi}{2}\sum_j \frac{\lambda_{j,n}^2}{Mm_j \omega_j^2} \delta(\omega-\omega_j)$
is the Fourier transform of the friction constant $\gamma_n(t)$
 (\ref{eq:EOM}) \cite{Hanggi}.
The memory damping can be also expressed in terms of the reservoir's spectral function
$g_n(\omega)= 2\pi \sum_j \frac{\lambda_{j,n}^2}{Mm_j \omega_j} \delta(\omega-\omega_j)$
as $\gamma_n(\omega)=g_n(\omega)/4\omega$.
Equation (\ref{eq:JH}) clearly manifests that in the harmonic limit the heat current is 
exclusively determined by the spectral properties of the reservoirs. 
This behavior results from the lack of
scattering mechanisms in the harmonic system.

As mentioned above, within this simple picture we can also model 
a highly anharmonic molecule by truncating the single mode spectrum.
For a two-level model ($l$=0,1) at high temperatures the heat current is given by \cite{Rectif}
\bea
J_A=\frac{\gamma_L(\omega_0) \gamma_R(\omega_0)} {\gamma_L(\omega_0)+\gamma_R(\omega_0)}
\frac{ \omega_0}{2T_{B}} \Delta T,
\label{eq:JA}
\eea
where the temperature of the local model (sometimes referred to as a bridge $B$) is 
\bea
T_B= \frac{\gamma_L(\omega_0) T_L + \gamma_R(\omega_0) T_R}
{\gamma_L(\omega_0)+\gamma_R(\omega_0)}.
\label{eq:TB}
\eea
Though expressions (\ref{eq:JH}) and (\ref{eq:JA}) provide the heat current for
a simplistic model,
they  may  serve us for gaining qualitative understanding of heat transfer in an $N$-sites 
molecule.
The key element here is the observation that in short linear chains relatively few modes
contribute to the transport of thermal energy \cite{heatcond}. For simplicity,
we may assume that a {\it single mode} dominates the dynamics,
and use the following generic form to describe its size dependence
\bea
\omega_0\approx \omega_M\left(1+\frac{\beta}{N}\right).
\label{eq:w0}
\eea
$\omega_M$ is the asymptotic frequency for large $N$, $\beta$ is a constant that is 
specific for the material.
Note that Eq. (\ref{eq:w0}) does not necessarily
describe the variation of the lowest vibrational mode of the
chain with increasing length. Rather, this is a qualitative expression
for the variation of the {\it dominant} heat conducting mode with size:
For short chains the spectrum is significantly altered with size.
For large enough chains $\omega_0$ is approximately fixed.

We can clearly discern in Eqs. (\ref{eq:JH})-(\ref{eq:JA}) the role of different factors 
(contacts  and internal interactions)
in determining the heat current. While the reservoirs spectral properties enter these expressions
through the damping rate $\gamma$, calculated
at the relevant molecular frequency $\omega_0$,  anharmonic interactions yield an extra $\omega_0/T_B$ factor
that accounts for the local mode occupancy. Since both of these terms depend on 
frequency, thus on size through
Eq. (\ref{eq:w0}), the resulting $N$ dependence is not trivial.
We study next various models for the damping element, Markovian
and non-Markovian, assuming for simplicity that the system is
symmetric with respect to the two ends, $\gamma=\gamma_n$ ($n=L,R$).

%
\begin{figure}[htbp]
\vspace{0mm} \hspace{0mm}
 {\hbox{\epsfxsize=80mm \epsffile{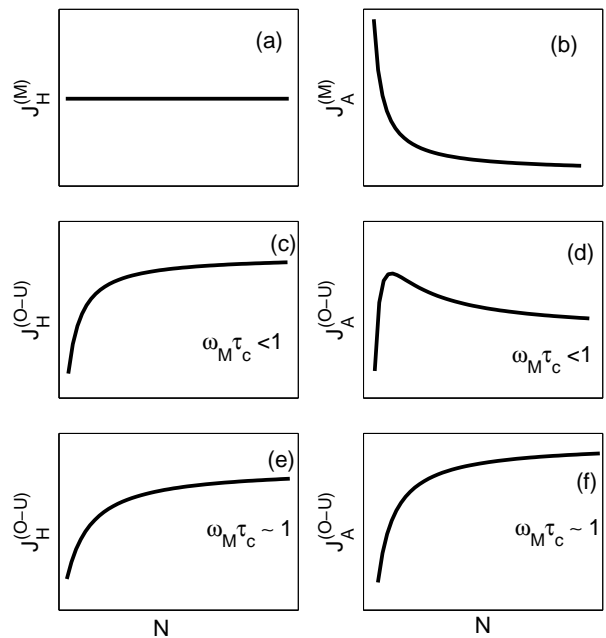}}}
\caption{
Qualitative behavior of the heat current in the single-mode heat conduction model
for Markovian and non-Markovian reservoirs.
(a)-(b) Heat current for a Gaussian white noise  [Eq. (\ref{eq:JM})];
(c)-(d) Current for an O-U noise with a short memory time, Eq. (\ref{eq:JOU}) with $\omega_M \tau_c < 1$;
(e)-(f) Current for an O-U noise with a long memory time, Eq. (\ref{eq:JOU}) with $\omega_M \tau_c \sim 1$.
}
\label{Fig1}
\end{figure}

\begin{figure}[htbp]
\vspace{0mm} \hspace{0mm}
{\hbox{\epsfxsize=80mm \epsffile{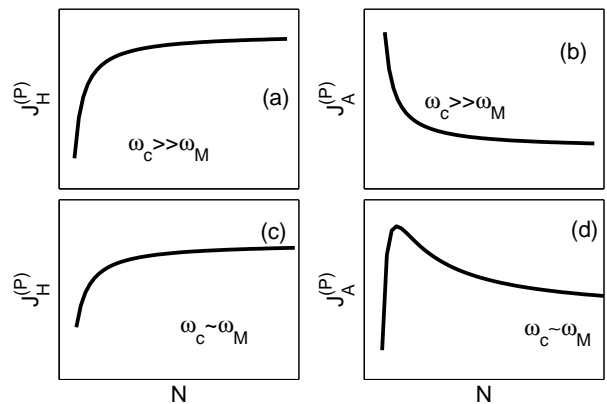}}}
\caption{
Qualitative behavior of heat current in the single-mode heat conduction model
with a power law spectral function (\ref{eq:JP}), $s=1$.}
\label{Fig2}
\end{figure}


{\it I. Markovian process}.
For a Markovian process the spectral function is Ohmic, $g(\omega)=4 \epsilon \omega$, thus
the friction is a constant, $\gamma(\omega)=\epsilon$.
In this limit Eqs. (\ref{eq:JH}) and (\ref{eq:JA}) reduce to
\bea
J_H^{(M)}&=&\frac{\epsilon}{2}k_B\Delta T
\nonumber\\
J_A^{(M)}&=&\frac{\epsilon\Delta T \omega_M}{2(T_L+T_R)} \left(1+\frac{\beta}{N}\right).
\label{eq:JM}
\eea
These expressions are consistent with standard results \cite{Lepri-rev}:
The current in harmonic systems does not depend on size,
while in anharmonic models it decays with distance.
In Fig. \ref{Fig1} (a)-(b) we qualitatively exemplify
this behavior by simulating Eq. (\ref{eq:JM}), reproducing the results of panels (a)-(b) of
Fig. \ref{Fig4}.

{\it II. Ornstein-Uhlenbeck noise}.
The exponentially correlated (O-U) process $\gamma(t)=\epsilon/\tau_c e^{-|t|/\tau_c}$ transforms
into a Lorentzian in frequency domain, $\gamma(\omega)=\frac{2\epsilon}{1+\omega^2\tau_c^2}$.
The fluxes (\ref{eq:JH}) and (\ref{eq:JA}) then become
($\epsilon=\epsilon_n$ and $\tau_c=\tau_c^n$)
\bea
J_H^{(O-U)} &=& \frac{k_B\Delta T \epsilon}{1+\omega_0^2 \tau_c^2} \propto  \left( 1-\frac{\beta}{N}\right)^2
\nonumber\\
J_A^{(O-U)}&=& \frac{\Delta T}{T_L+T_R} \frac{\epsilon \omega_0}{1+\omega_0^2 \tau_c^2}
\propto \left(1-\frac{\beta}{N}\right),
\label{eq:JOU}
\eea
where we derived the approximate distance dependence
by assuming that $\omega_0\tau_c>1$ and taking $\beta/N<1$.
These expressions clearly demonstrate an {\it enhancement} of the current
with $N$ for both harmonic and anharmonic
systems when the reservoirs have (each) a highly correlated noise.

In Fig. \ref{Fig1} (c)-(f) we simulate Eq. (\ref{eq:JOU}).
For short correlation time we reproduce the results of Fig. \ref{Fig4}(b),
manifesting an enhancement of the anharmonic flux followed by a decrease of current.
For very long correlation times the current systematically increases with size,
in agreement with the numerical data, Fig. \ref{Fig4} (d).


{\it III. Power law models}.
The spectral function $g(\omega)=\epsilon \omega^{s}e^{-\omega/\omega_c}$ 
is a widely accepted description of solids, 
$\omega_c$ is the reservoir cutoff frequency.
Unlike the O-U noise which satisfies the differential equations (\ref{eq:EOM2}),
this noise  cannot be  reduced into a multi-component Markovian process,
thus it is not trivial to simulate \cite{Fox}.
Within the single-mode heat conduction model
the heat current for harmonic and anharmonic modes (\ref{eq:JH})-(\ref{eq:JA}) is given by
\bea
J_H^{(P)}&=&\frac{\epsilon}{2}k_B \Delta T \omega_0^{s-1} e^{-\omega_0/\omega_c}
\nonumber\\
J_A^{(P)}&=&\epsilon\frac{\Delta T}{2(T_L+T_R)}\omega_0^{s} e^{-\omega_0/\omega_c},
\label{eq:JP}
\eea
where $\omega_0$ is the dominant frequency for heat transport (\ref{eq:w0}).
For an Ohmic ($s=1$) bath, taking $\omega_c\gg \omega_0$, the dynamics
reduces to the Markovian limit (\ref{eq:JM}), where $J_H^{(P)}$ is weakly
enhanced with distance and $J_A^{(P)}$ decays with $N$.
This behavior is exemplified in Fig. \ref{Fig2} (a)-(b).
In contrast, for $\omega_c \sim \omega_0$, a different trend is observed:
while $J_H^{(P)}$ monotonically increases with length (c), in anharmonic
systems the current first increases with size, then falls down (d).
For $s=2$ we expect that, quite interestingly, the heat current in harmonic models will
{\it decay} with distance. This is because 
for long chains the dominant conducting frequencies
are red shifted, while the solid spectral function is maximal at $\omega_c$.

We conclude that short to intermediate size molecular chains coupled to general environments
can manifest a rich behavior:
The thermal current can either increase or decrease with
size, critically depending both on the molecular internal interactions and on the reservoirs
spectral properties.
Despite its simplicity, the model presented here provides a  useful starting point
for explaining the complicated behavior observed within atomistic 
classical molecular dynamics simulations.
In the next section we show that this model is also a useful tool for estimating the
temperature profile along molecule.

\section{Temperature profile}

We investigate next the effect of the reservoirs spectral properties on the
local temperature profile in harmonic and anharmonic chains. Within the
single mode heat conduction model one can
define in steady state  the temperature  of the local mode
as $k_B T_B=\omega_0\sum_l l P_l$,  where $P_l$ is the population
of the $l$ vibrational state. In the classical limit it can be
shown that this expression reduces to Eq. (\ref{eq:TB}),
$ T_{B}= \frac{\gamma_L(\omega_0) T_L+\gamma_R(\omega_0) T_R} {\gamma_L(\omega_0)+\gamma_R(\omega_0)}$,
for both harmonic and anharmonic local modes \cite{Rectif, NDR}. 
We therefore expect that when the reservoirs have the same type of spectral function with the same coefficients,
the temperature profile is independent of the type, and is given by the arithmetic average
$T_{B}=(T_L+T_R)/2$. On the other hand, when the two reservoirs
have different types of spectral densities (or different memory times),
the temperature profile along the chain depends on the model.
For example, if $T_L\ll
T_R$ and  $\gamma_R\ll\gamma_L$,  $T_{B} \sim \gamma_R
T_R/\gamma_L \ll T_R$. In contrast, for $T_L\ll T_R$ but $\gamma_R>\gamma_L$,
$T_{B}\sim T_R$.
This behavior was observed by Saito et al. in the harmonic limit
using a quantum master equation formalism \cite{Saito}.

Using classical molecular dynamics simulations
we compute the temperature profile by calculating the mean kinetic energy for each particle,
 $T_k=\langle \frac{p_k^2}{2m} \rangle \frac{2}{k_B}$.
Here $p_k$ is the momentum of the $k$th particle  calculated after steady state develops, 
and we average over time and initial
configurations using the procedure and parameters of section II.

\begin{figure}[htbp]
\vspace{0mm} \hspace{0mm}
{\hbox{\epsfxsize=85mm \epsffile{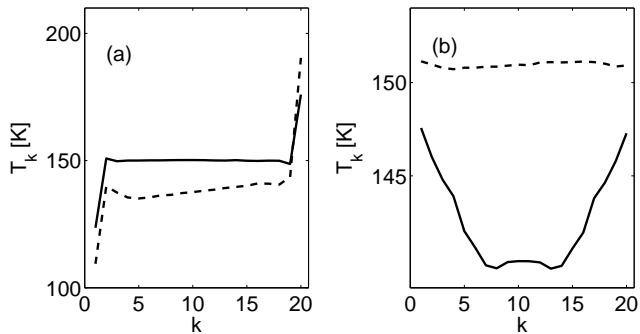}}}
\caption{Temperature profile along harmonic (full) and anharmonic (dashed)  chains
coupled to reservoirs of identical spectral properties.
(a) Gaussian white noise; (b) O-U noise with $\tau_c=0.04$ ps.
$N$=20, $T_R$=300 K, $T_L$=0 K and $\epsilon=50$ ps$^{-1}$ in all cases.
}
\label{Tg1}
\end{figure}

Fig. \ref{Tg1} manifests that the bulk temperature profile 
is close to the arithmetic average (150 K) for both Markovian and non-Markovian
systems when the reservoirs are identical.
At the boundaries we see interesting features noted earlier in Refs. 
\cite{Lebowitz,Saito, Zurcher,DharPRB}:
The temperature close to the cold end ($L$)  slightly rises, even to values higher than the average
temperature.
For an anharmonic system a small temperature gradient develops at the chain center.
This effect is more pronounced for Markovian reservoirs. 
Note that in Markovian harmonic systems 
the temperature at the chain center, e.g. $T_{10}$, 
is higher than the anharmonic value.
This trend is reversed for O-U baths, as here anharmonic interactions facilitate
equilibration with the slow environment.

\begin{figure}[htbp]
\vspace{0mm} \hspace{0mm}
{\hbox{\epsfxsize=70mm \epsffile{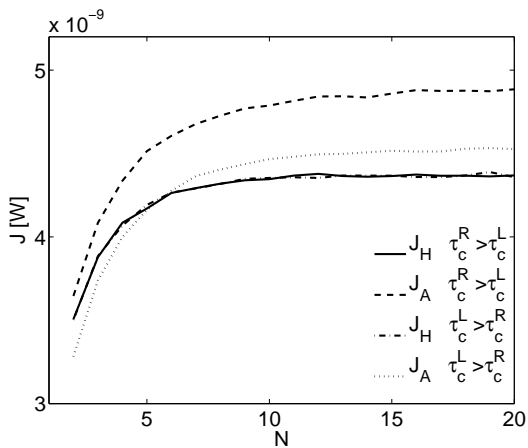}}}
\caption{Heat current in molecular chains coupled to reservoirs
of different spectral properties $\tau_c^L \neq \tau_c^{R}$, $\epsilon_L=\epsilon_R=$50 ps$^{-1}$.
$\tau_c^L=0.002$ ps, $\tau_c^R=0.04$ ps: harmonic (full);  anharmonic  (dashed).
$\tau_c^L=0.04$ ps, $\tau_c^R=0.002$ ps: harmonic (dashed-dotted); anharmonic (dotted).
$T_R$=300 K and $T_L$=0 K in all cases.
}
\label{Tg2}
\end{figure}

\begin{figure}[htbp]
\vspace{4mm} \hspace{0mm}
{\hbox{\epsfxsize=80mm \epsffile{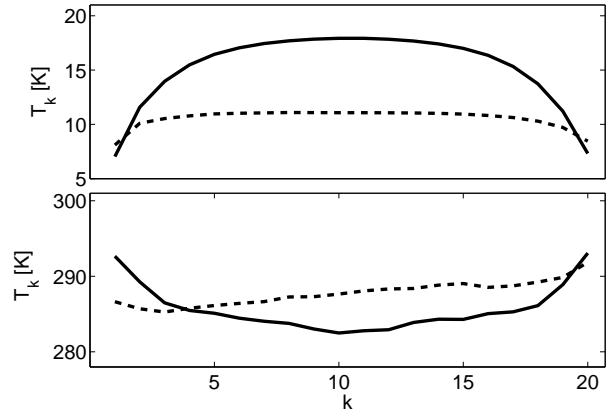}}}
\caption{Temperature profile along harmonic (full) and anharmonic (dashed) chains
coupled to reservoirs of different spectral properties
$\tau_c^L \neq \tau_c^{R}$, $\epsilon_L=\epsilon_R=$50 ps$^{-1}$.
$\tau_c^R=0.04$ ps, $\tau_c^L=0.002$ ps (top);
$\tau_c^R=0.002$ ps, $\tau_c^L=0.04$ ps (bottom).
The temperature profile was calculated for an $N=20$ sites chain using
$T_R$=300 K, $T_L$=0 K in all cases.
}
\label{Tg3}
\end{figure}

We employ next reservoirs of different spectral properties, and study the heat current and
temperature profile in this asymmetric system.
We assume that both reservoirs are of O-U type, but take different
memory times $\tau_c^L\neq \tau_c^R$.
Fig. \ref{Tg2} manifests that the current in a harmonic system is independent
of the direction of the asymmetry, i.e. if the cold bath is weakly or strongly coupled to the
molecule. In contrast, the current in an  anharmonic chain is
significantly  altered when the asymmetry is reversed: It is higher when
the hot bath is slow.
This response to spatial asymmetry is the underlying principle of the "thermal rectifier" 
\cite{RectifE,Rectif,Terraneo}.

Fig. \ref{Tg3} shows that
the bridge temperature dramatically responds to the direction of the asymmetry
for {\it both harmonic and anharmonic systems}. We find that it is close to the temperature of the
reservoir with the short memory time. In other words, the
molecule better equilibrates with the Markovian reservoir.
This behavior is consistent with the results of Ref. \cite{Saito},
and qualitatively agrees with Eq. (\ref{eq:TB}).




\section{Discussion and Summary}

In this paper we have investigated the effect of the contacts' spectral properties
on the thermal conduction of harmonic and anharmonic short to intermediate 
size chains using classical molecular dynamics simulations.

When the system size is  smaller than the mean free path,
heat conduction is dominated by harmonic interactions. 
In such systems the spectral properties of the contacts play a central role 
in determining the dynamics in the junction.
%
We found that the distance dependence of the current strongly varies with the baths (noise) correlation time:
While for Markovian baths the current in harmonic systems is independent of size, it slightly
increases with length when the reservoirs have long memory. The effect 
is even more dramatic in anharmonic chains. Here the current decays with $N$ in the Markovian limit,
while it gets enhanced with distance when the reservoirs have long memory times.

Other interesting observations are: (i) We identified a regime where the anharmonic flux
is practically distance independent due to the counteracting effects
of molecular nonlinearities and bath correlations.
(ii) For long memory times the current through anharmonic chains might be
higher than the analogous harmonic current. 
Thus, anharmonic interactions might play a surprising role, 
enhancing the heat flux across molecules
coupled to non-Markovian reservoirs.  
We also showed that
the single-mode heat conduction model developed in Ref. \cite{Rectif}
can essentially capture the length dependence and the temperature profile obtained within classical
molecular dynamics simulations.

Before we conclude we briefly discuss the relationship of our calculations to relevant 
experiments.
Wang et al. have recently measured the thermal
conductance ($\mathcal {K} =J/\mathcal{A} \Delta T $, $\mathcal {A}$ is the cross section)
of 8-10 alkandithiols monolayers \cite{Segalman}. 
The values $\mathcal {K}\sim$ 25-28 MWm$^{-2}$K$^{-1}$ were obtained, 
roughly independent of system size.
Previous calculations \cite{heatcond} confirm this observation:
The heat current of alkane chains is expected to be distance
independent, even for very long ($N$=100) molecules, due to weak anharmonic internal interactions in the system.
More quantitatively, Eq. (\ref{eq:JH}) predicts thermal conductance of
$\mathcal {K} \sim 280$ MWm$^{-2}$K$^{-1}$, using Markovian
thermal baths with $\epsilon$=10 ps$^{-1}$ 
and a cross section $\mathcal{A}=(5\times10^{-10})^2$m$^{-2}$.
As discussed above, memory effects can reduce this number by 1-2 orders of magnitude.
Therefore the conductance calculated here is in a reasonable agreement with experimental results.

Several future directions are of interest.
First, we have restricted our discussion to the O-U correlation function,
since it can be easily simulated as a multi-component Markovian process \cite{Luczka}.
It is worth extending this study to include more realistic
environments of $\omega^{s}$ spectral properties.
This type of noise correlations can be simulated using the inverse Fourier
transform technique \cite{Fox,Bao,Baoheat}.
Other interesting questions are what is the role of the contacts in higher dimensions, and
how do quantum mechanical effects alter
the phenomena addressed in this paper \cite{Saito,heatcond,Zurcher, DharPRB,Cirac,Mingo,Wang}.

Understanding thermal transport in molecule-solid interfaces is important for
molecular electronic applications \cite{MolEl}. It is also an imperative step
in the endeavor for developing unique molecular level thermal devices \cite{Gate,Circuit,pump}.


\begin{acknowledgments}
This work was supported by a University of Toronto Start-up Grant.
\end{acknowledgments}




\begin{thebibliography}{32}

\bibitem{Kim}
P.  Kim , L. Shi, A.  Majumdar, P. L. McEuen, Phys. Rev. Lett. {\bf
87}, 215502 (2001).

\bibitem{Braun}
Z. Ge, D. G. Cahill, P. V. Braun, Phys. Rev. Lett. {\bf 96},
186101  (2006).

\bibitem{Segalman}
R. Y. Wang, R. A. Segalman, A. Majumdar,
App. Phys. Lett. {\bf 89}, 173113 (2006).

\bibitem{RectifE}
C. W. Chang, D. Okawa, A. Majumdar, A. Zettl, Science {\bf 314},
1121 (2006).

\bibitem{Dlott}
Z. Wang, J. A. Carter, A. Lagutchev, Y. K. Koh, N.-H. Seong, 
D. G. Cahill, D. D. Dlott,
Science, {\bf 317}, 787 (2007).

\bibitem{films}
C. Chiritescu, D. G. Cahill, N. Nguyen, D. Johnson, A. Bodapati, 
P. Keblinski, P. Zschack,
Science {\bf 315}, 351 (2007).

\bibitem{Tunable}
C. W. Chang, D. Okawa, H. Garcia, T. D. Yuzvinsky, A. Majumdar, A. Zettl,
App. Phys. Lett. {\bf 90}, 193114 (2007).

\bibitem{pump}
D. Segal, A. Nitzan, Phys. Rev. E {\bf 73}, 026109 (2006).

\bibitem{Gate}
L. Wang, B. Li,
Phys. Rev. Lett. {\bf 99}, 177208 (2007).

\bibitem{Circuit}
Z. Liu, B. Li, Phys. Rev. E {\bf 76}, 051118 (2007).


\bibitem{Reddy}
P. Reddy, S.-Y. Jang, R. A. Segalman, A. Majumdar,
Science  {\bf 315}, 1568  (2007).

\bibitem{Hochbaum}
A. I. Hochbaum, R. K. Chen, R. D. Delgado, W. J. Liang, E. C. Garnett, 
M. Najarian, A. Majumdar, P. D. Yang, 
Nature {\bf 451}, 163 (2008).

\bibitem{Lepri-rev}
S. Lepri, R. Livi, A. Politi, Phys. Rep. {\bf 377}, 1 (2003).

\bibitem{Lebowitz}
Z. Rieder, J. L. Lebowitz, E. Lieb, J. Math. Phys. {\bf 8}, 1073 (1967).

\bibitem{Dhar}
A. Dhar, Phys. Rev. Lett. {\bf 86}, 5882 (2001).


\bibitem{Saito}
K. Saito, S. Takesue, S. Miyashita, Phys. Rev. E {\bf 61}, 2397 (2000).

\bibitem{Dhar2D}
L. Wee Lee, A. Dhar, Phys. Rev. Lett. {\bf 95}, 094302 (2005).

\bibitem{Zhao}
H. Zhao, L. Yi, F. Liu, B. Xu,
Eur. Phys. J. B {\bf 54}, 185 (2006).

\bibitem{Barik}
D. Barik, Eur. Phys. J. B {\bf 56}, 229 (2007).

\bibitem{Rectif}
D. Segal, A. Nitzan, Phys. Rev. Lett.  {\bf 94}, 034301 (2005);
 J. Chem. Phys. {\bf 122}, 194704  (2005).

\bibitem{NDR}
D. Segal, Phys. Rev. B {\bf 73}, 205415 (2006).

\bibitem{OU}
G. E. Uhlenbeck, L. S. Ornstein, Phys. Rev. {\bf 36}, 823 (1930).

\bibitem{Luczka}
J. Luczka, Chaos {\bf 15}, 026107 (2005).

\bibitem{Lifson}
S. Lifson, P. S. Stern, J. Chem. Phys. {\bf 77}, 4542 (1982).

\bibitem{heatcond}
D. Segal, A Nitzan, P. H\"anggi, J. Chem. Phys. {\bf 119}, 6840
(2003).

\bibitem{Hanggi}
P. H\"anggi, Lect. Notes Phys. {\bf 484}, 15 (1997).

\bibitem{Fox}
R. F. Fox, I. R. Gatland, R. Roy, G. Vemuri,
Phys. Rev. A {\bf 38}, 5938 (1988).

\bibitem{Zurcher}
U. Z\"urcher,  P. Talkner, Phys. Rev. A {\bf 42}, 3278 (1990).

\bibitem{DharPRB}
A. Dhar, B. S. Shastry, Phys. Rev. B {\bf 67}, 195405 (2003).

\bibitem{Terraneo}
M. Terraneo, M. Peyrard,  G. Casati,
Phys. Rev. Lett. {\bf 88}, 094302 (2002).


\bibitem{Bao}
J.-D. Bao, Y.-Z. Zhuo,
Phys. Rev. E {\bf 71}, 010102(R) (2005).

\bibitem{Baoheat}
X.-P. Zhang, J.-D. Bao, Phys. Rev. E {\bf 73}, 061103 (2006).

\bibitem{Cirac}
A. Ozpineci, S. Ciraci, Phys. Rev. B {\bf 63}, 125415 (2001).

\bibitem{Mingo}
N. Mingo,
Phys. Rev. B {\bf 74}, 125402 (2006).

\bibitem{Wang}
J.-S. Wang,
Phys. Rev. Lett. {\bf 99}, 160601 (2007).

\bibitem{MolEl}
M. Galperin, A. Nitzan, M. A. Ratner, 
Phys. Rev. B {\bf 75}, 155312 (2007).


\end{thebibliography}
\end{document}